\documentclass[english,a4paper,14pt]{article}
\usepackage{babel,amssymb,graphicx,hyperref}

\advance\textheight \topskip \textwidth 35.5pc \oddsidemargin 20pt
\begin{document}

\author{V. N. Lukash\footnote{E-mail:lukash@asc.rssi.ru}}
\title{\bf{On the relation between tensor and scalar
perturbation modes in Friedmann cosmology}\footnote{An extended
version of the paper published in Usp. Fiz. Nauk {\bf176} 113
(2006) [Phys. Usp. {\bf 49} 103 (2006)].}}
\date{\emph{Astro Space Centre of the P.N.Lebedev Physics Institute, Moscow}}
\maketitle
\begin{abstract}
An elementary derivation of the fundamental relation $T/S=4\gamma$
between the tensor and scalar modes of cosmological perturbations
in the early universe is given. Statements by L.P.Grishchuk on
this problem are commented on.
\end{abstract}

\section{Motivation}

In a recent paper ``Relic gravitational waves and cosmology'' by
L.P.Grishchuk [see Phys. Usp. \textbf{48} (12) 1235 (2005)],
author has reviewed his studies conducted over many years and
devoted to quantum-gravity generation of gravitational waves
(tensor mode $T$) and density perturbations (scalar mode $S$) in a
homogeneous isotropic universe. The origin of primordial
cosmological perturbations has been a key question of 20th century
physics, initiated by the pioneering work by E.M.Lifshitz
\cite{Lifshitz} and the first papers where quantization of $T$
\cite{Grishuk-2} and $S$ \cite{Lukash} modes of perturbations in a
flat Friedmann model has been done. The view by the author of Ref.
\cite{Grishuk} in its principal points contradicts the widely
accepted result recognized as classical on the $T$-to-$S$ mode
ratio in the early universe, which is included in textbooks on
cosmology.

Inasmuch as the central points of Ref. \cite{Grishuk} and some
earlier papers by Grishchuk, devoted to the relation between
spectra of relic gravitational waves and density perturbations,
are based on the statement that the ``\textit{final amplitudes of
gravitational waves and density perturbations should be roughly
equal to each other''} [see the discussion after
formula\footnote{Hereinafter references to formulas from paper
\cite{Grishuk} are marked with an asterisk.}
$(33)^{\displaystyle\ast}$], in the present paper we shall
consider only this key statement.

The cited statement runs counter to the generally accepted result
that the \textit{ratio of the squares of the amplitudes of the
$T$- and $S$-modes of cosmological perturbations generated
quantum-gravitationally in the early universe is proportional to
$\gamma$}, where the parameter $\gamma\equiv -\dot{H}/H^{2}$ is
taken at the initial time of the parametric amplification for
perturbations of a given wavelength\footnote{In the exact theory,
the result will also depend on another parameter $\beta$, the
speed of sound in the medium in units of the speed of light
\cite{Lukash}. However, in most applications considered $\beta\sim
1$, so in what follows we omit the parameter $\beta$. We also use
units where $c=1$ and $l_{\rm{Pl}}=(G\hbar)^{1/2}$, the scale
factor is $a\equiv(1+z)^{-1}$, $\mathbf{x}$ are the spatial
(comoving) coordinates in the flat Friedmann model, $n$ is the
Fourrier wavenumber, $\eta=\int dt/a$ and $t$ are the conformal
and physical time, respectively, $H=\dot{a}/a=a'/a^{2}$ is the
Hubble parameter, and a dot or prime over a function means its
derivative with respect to the physical or conformal time,
respectively. We shall mark with a subscript $T$ or $S$ any
variable while considering $T$ and $S$ perturbation modes,
respectively.}. (We are reminded that at the inflationary stage
$\gamma<1$.) Below, we shall recall the importance of $T$ and $S$
for the modern cosmology, present the elementary derivation of the
classical relationship $T/S=4\gamma$ without solving the equation
for perturbations, and then show what is erroneous in the
statements by Grishchuk.

\section{T \& S, the witnesses of Big Bang}

The geometry of our universe is fully described by four functions:
the scale factor of spatial expansion, $a(t)$, and three spectral
functions, $S(n), V(n), T(n)$ determining the scalar, vector and
tensor Gaussian modes of the metric tensor at large scale ($n <
aH\equiv \dot{a}$). In the course of Friedmannian expansion scales
enter the horizon, $V$ and $T$ modes decay in time leaving
imprints in CMB anisotropy and polarization, but $S$ mode grows
due to gravitational instability and creates the large scale
structure of observable universe. $V$ mode is not currently
considered since the lack of relevant sources. $S$ and $T$ modes
are fundamental, their discovery is a hot topic of the modern
cosmology as they sufficiently determine the model of the universe
at period of its creation. $S$ mode is found up to accuracy of
$\sim 10\%$, the detection of $T$ mode is a matter of nearest
future. $T$ mode is straightforwardly related to the Hubble scale
of the very early Universe, the ratio $T/S$ is linked to the time
derivative of $H$. Thus, when both known, we discover the model of
the universe at period of $T$ and $S$ generation.

\section{Initial conditions}

From the theoretical point of view, the problem of small linear
cosmological perturbations is equivalent to the problem of the
behavior of test fields in an unperturbed Friedmann model which is
reduced in turn to the problem of massless real fields in the
Minkowski space-time, evolving under the influence of an external
variable field:
\begin{equation}
\label{formula-1} S[q]=\int Ld\eta
d\mathbf{x},~~~L=\frac{1}{2}\alpha^{2}\eta_{\mu\nu}q_{,\mu}q_{,\nu},
\end{equation}
where $S$ and $L$ are the action and the Lagrangian density of the
field $q$, respectively, and the comma in the subscript stands for
the derivative over the Minkowskian coordinates
$(\eta,\mathbf{x})$ with the metric tensor
${\eta_{\mu\nu}=(1,-1,-1,-1)}$.

The role of the external (parametric) field is played by the
time-dependent function $\alpha^{2}$. It is equal to
$\alpha^{2}_{T}=a^{2}/8\pi G$ for each of two polarizations of
gravitational waves (in this case, $q_{T}$ is the
transverse-traceless component of the gravitational
field)~\cite{Grishuk-2}, and is $\alpha^{2}_{S}=a^{2}\gamma/4\pi
G$ for density perturbations (in that case, $q_{S}$ is the
gauge-invariant combination of the longitudinal gravitational
potential which is perturbation of the scale factor, and the
potential of the 3-velocity of the medium multiplied by the Hubble
parameter)~\cite{Lukash}.

In the Fourier representation, the field $q$ is resolved into
elementary time-dependent oscillators $q_{n}$ with the Lagrangians
(below we shall omit the subscript $n$ of the Fourier modes)
\begin{equation}
\label{formula-2}
L_{n}=\frac{1}{2n^{3}}\alpha^{2}(q'^{2}-n^{2}q^{2}).
\end{equation}

The evolution of oscillator~(\ref{formula-2}) depends on the
function $f$ determining its effective frequency:
\begin{equation}
\label{formula-3}
\begin{array}{l}
\bar{q}''+n^{2}(1-f)\bar{q}=0,\\
\bar{q}=\displaystyle\frac{\alpha}{n}q,~~~f\equiv\displaystyle\frac{\alpha''}{\alpha
n^{2}}.
\end{array}
\end{equation}
When $|f|\ll 1$, the oscillator $q$ stays in the free adiabatic
oscillation regime and decays inversely proportional to $\alpha$
($q\propto\exp{(-in\eta)}/\alpha$). When $f\geq 1$, a parametric
amplification occurs and the field $q$ `freezes out' ($q\propto
const$). In variables $(\bar{q},\bar{p})$, the Lagrangian takes
the standard canonical form
\begin{equation}
\label{formula-4}
\begin{array}{lc}
L_{n}=\displaystyle\frac{n}{2}(\bar{p}^{2}-\bar{q}^{2}),&~\\
~&~\\
\bar{p}=\displaystyle\frac{\partial
L_{n}}{\partial\bar{q}'}=\frac{\alpha
q'}{n^{2}}=\frac{\bar{q}'}{n}-s\bar{q},&
s\equiv\displaystyle\frac{\alpha'}{\alpha n},
\end{array}
\end{equation}
where $\bar{p}$ is the field momentum conjugate to $\bar{q}$.

The key to understanding the $T/S$ ratio lies in choosing the
initial conditions for elementary $q$-oscillators. It is
convenient to determine them in the adiabatic zone as states with
minimum energy for all $T$- and $S$-oscillators (vacuum). The
quantization of systems~(\ref{formula-1}) and~(\ref{formula-2}) is
a standard procedure that does not require explanation. The
question is in an unambiguous choice of the initial vacuum state
for the $q$-oscillator. We are reminded that a free
(noninteracting) oscillator possesses a unique ground state.

\section{T/S for taxpayers}

In most scenarios of the early universe, the adiabatic condition
is realized in a microscopic region ($\eta<\eta_{i}$), where the
period of oscillations of the $q$-oscillator is smaller than the
characteristic variability time of the parameter~$\alpha$:
\begin{equation}
\label{formula-5} |s|<1,~~|f|<1,
\end{equation}
and the Hamiltonian of system~(\ref{formula-2}) is positively
determined. When both conditions~(\ref{formula-5}) are satisfied,
quantum-mechanical operators $\bar{q}$ and $\bar{p}$ describe in
the leading order an oscillator free of external action [see
Eqns~(\ref{formula-3}) and (\ref{formula-4})]. This allows us to
use the standard procedure for the frequency decomposition (into
positive and negative sets) and for the determination of the
ground state [the absence of particles at
stage~(\ref{formula-5})]:
\begin{equation}
\label{formula-6}
\langle\bar{p}^{2}\rangle=\langle\bar{q}^{2}\rangle=\frac{\hbar}{2},
~~\eta<\eta_{i},
\end{equation}
where brackets $\langle\ldots\rangle$ signify averaging over the
given (vacuum) state.

Equivalently, the ground state at stage (\ref{formula-5}) can be
constructed using the `normalized' variables~$q$:
\begin{equation}
\label{formula-7}
\begin{array}{l}
\tilde{q}=\displaystyle\frac{\alpha_{0}}{n}q,\\
~\\
L_{n}=\displaystyle\frac{1}{2n}\bar{\alpha}^{2}({\tilde{q}}'^{2}
-n^{2}\tilde{q}^{2})=\frac{n}{2}\left(\frac{\tilde{p}^{2}}{\bar{
\alpha}^{2}}-\bar{\alpha}^{2}\tilde{q}^{2}\right),\\
~\\
\bar{\alpha}\equiv\displaystyle\frac{\alpha}{\alpha_{0}},
\end{array}
\end{equation}
which canonize the Lagrangian within some period of time at any
instant $\eta_{0}$ (with $\eta_{0}<\eta_{i}$), within which the
value of $\alpha$ can be considered constant ($\bar{\alpha}\simeq
1$). Therefore, canonical pairs $\tilde{q}\simeq\bar{q}$ and
$\tilde{p}\simeq\bar{p}$ and vacuum conditions (\ref{formula-6}),
(\ref{formula-8}) turn out to be identical:
\begin{equation}
\label{formula-8}
\langle\tilde{p}^{2}\rangle=\langle\tilde{q}^{2}\rangle=\frac{\hbar}{2},
~~\forall~\eta_{0}<\eta_{i}.
\end{equation}
(One can say that variables $\tilde{q}$ form a tangent space to
the function $\bar{q}$).

We stress that the adiabatic conditions (\ref{formula-5}) provide
the unique choice of the state (\ref{formula-6}),
(\ref{formula-8}) as the initial state of elementary oscillators
(\ref{formula-2}) that corresponds to the minimal initial level of
their excitations (vacuum of the field $q$). In a later evolution,
$q$-oscillators enter the zone of the parametric amplification,
equalities (\ref{formula-6}) and (\ref{formula-8}) are violated,
and their state becomes multiparticle (the generation of
cosmological perturbations).

Assuming the existence of the adiabatic stage (\ref{formula-5}) in
the early universe, which at the instant $\eta_{i}$ ($f\sim s=1$)
changed to the parametric amplification stage, we obtain from
condition (\ref{formula-6}) the amplitude of the $q$-oscillator in
the `freezing-out' zone ($\eta>\eta_{i}$):
\begin{equation}
\label{formula-9} \langle q^{2}\rangle\simeq\langle
q_{i}^{2}\rangle=\frac{n^{2}}{\alpha_{i}^{2}}\langle
\bar{q}_{i}^{2}\rangle\approx\frac{\hbar n^{2}}{2\alpha_{i}^{2}}.
\end{equation}
Hence follows the validity of the generally recognized statement
for the ratio of the perturbation modes of a given
wavelength\footnote{Accounting for $\alpha_{i}H_{i}\approx n$,
from formulas (\ref{formula-9}) and (\ref{formula-10}) we obtain
the well- known expressions for the spectra of perturbations and
their slopes:
$$
\begin{array}{ll}
\langle q_{T}^{2}\rangle^{1/2}\approx l_{Pl}H_{i},&
n_{T}\equiv\displaystyle\frac{d\ln{\langle
q_{T}^{2}\rangle}}{d\ln{n}}\simeq
-2\gamma_{i}\simeq -0.5\frac{T}{S},\\
~\\
\langle q_{S}^{2}\rangle^{1/2}\approx
\displaystyle\frac{l_{Pl}H_{i}}{(2\gamma_{i})^{1/2}},&
n_{S}\equiv\displaystyle\frac{d\ln{\langle
q_{S}^{2}\rangle}}{d\ln{n}}\simeq - 2\gamma_i-
\frac{\dot{\gamma}}{\gamma H} \big\vert_i.
\end{array}
$$
}:
\begin{equation}
\label{formula-10} \frac{T}{S}\equiv \left.2\frac{\langle
q_{T}^{2}\rangle}{\langle
q_{S}^{2}\rangle}\right|_{\eta>\eta_{i}}\simeq
2\left(\frac{\alpha_{T}}{\alpha_{S}}\right)^{2}_{i}=4\gamma_{i}
\end{equation}
(both polarizations of gravitational waves were taken into
account).

\section{NB: about a mistake}

Now let us consider Grishchuk's error. The dimensional amplitudes
of elementary oscillators correspond to the following notations
from Ref. \cite{Grishuk}:
$$
q_{T}\equiv h,~q_{S}\equiv\frac{\zeta}{2}.
$$
[see formulas $(11)^{\displaystyle\ast}$,
$(20)^{\displaystyle\ast}$]. When determining the state of the
$T$-oscillator, the author follows equations (\ref{formula-7}) and
(\ref{formula-8}) ($\tilde q_T \equiv\bar h$); see Eqns
$(12)$-$(17)^{\displaystyle\ast}$). However, when moving to the
$S$-mode, instead of normalized variable $\tilde q_S$ he
introduces an asymmetric (with respect to $\alpha_{S}\propto
a\sqrt{\gamma}$) variable $\bar{\zeta}$ [see Eqn
$(21)^{\displaystyle\ast}$]:
\begin{equation}
\label{formula-11}
\begin{array}{l}
\bar{\zeta}\equiv q_{LPG}=\displaystyle\frac{\tilde{q}_{S}}{\sqrt{\gamma_{0}}},\\
~\\
L_{n}=\displaystyle\frac{1}{2n}\bar{\alpha}_{S}^{2}\gamma_{0}({\bar{\zeta}}'^{2}
-n^{2}\bar{\zeta}^{2})=\frac{n}{2}\left(\frac{p_{LPG}^{2}}{\bar{\alpha}_{S}^{2}
\gamma_{0}}-\bar{\alpha}_{S}^{2}\gamma_{0}q_{LPG}^{2}\right),\\
~\\
\bar{\alpha}_{S}\equiv\displaystyle\frac{a\sqrt{\gamma}}{a_{0}\sqrt{\gamma_{0}}},
\end{array}
\end{equation}
for which the Lagrangian explicitly depends on $\gamma_{0}$. Then
equations (\ref{formula-8}), rewritten for the pair
$$
q_{LPG}\equiv\bar{\zeta},~~p_{LPG}=\frac{\partial
L_{n}}{\partial\bar{\zeta}'}=\tilde{p}\sqrt{\gamma_{0}}
$$
[$\rm \bf q$, $p$ in notations of Eqns $(24)^{\displaystyle\ast}$,
$(25)^{\displaystyle\ast}$], also acquire the explicit dependence
on the parameter $\gamma$:
\begin{equation}
\label{formula-12} \langle
q_{LPG}^{2}\rangle=\frac{\hbar}{2\gamma_{0}},~~\langle
p_{LPG}^{2}\rangle=\frac{\hbar}{2}\gamma_{0},~~\forall~\eta_{0}<\eta_{i}.
\end{equation}

Clearly, the vacuum state is in no way related to the choice of
one pair of canonical variables or another. Equations
(\ref{formula-8}) [and identical to them Eqns (\ref{formula-12})]
bears a transparent invariant sense: the equality of quantities
$\langle \tilde{q}^{2}\rangle=\langle \tilde{p}^{2}\rangle$ (or
$\gamma_{0}\langle q_{LPG}^{2}\rangle=\langle
p_{LPG}^{2}\rangle/\gamma_{0}$) means the equality of the mean
kinetic and potential energies of the elementary oscillator
(\ref{formula-2}), while the equality of each of these quantities
to $\hbar/2$ means choosing the minimum possible energy level of
the oscillator (i.e., the vacuum state) at the adiabatic stage
(\ref{formula-5}).

Nevertheless, the author of Ref. \cite{Grishuk} erroneously
interprets the state (\ref{formula-12}) for $S$-oscillators as a
squeezed one [multiparticle; see formulas after Eqns
$(26)^{\displaystyle\ast}$, $(27)^{\displaystyle\ast}$], ignoring
the fact that the asymmetry of equations (\ref{formula-12}) has
nothing to do with the choice of the state over which the
averaging was performed, but to the choice of the
variable\footnote{We recall that $S$-oscillators are coupled to
the product $a\sqrt{\gamma}$ and not to $a$ or $\gamma$
separately.} that is explicitly dependent on $\gamma$. This leads
him to introduce another initial state (we shall mark this state
by the subscript LPG) which he calls ``\textit{the genuine vacuum
state for the variable $\zeta$}'' [see formulas after Eqn
$(32)^{\displaystyle\ast}$]:
\begin{equation}
\label{formula-13} \langle q_{LPG}^{2}\rangle_{LPG}=\langle
p_{LPG}^{2}\rangle_{LPG}=\frac{\hbar}{2},~~\forall~\eta_{0}<\eta_{i},
\end{equation}
and, as a consequence, to the statement that $T/S\sim 1$ [see Eqn
$(33)^{\displaystyle\ast}$], since for $\eta>\eta_{i}$ one finds
\begin{equation}
\label{formula-14} \langle q_{S}^{2}\rangle_{LPG}\simeq
\gamma_{i}\langle
q_{S}^{2}\rangle\approx\left(\frac{l_{Pl}n}{a_{i}}\right)^{2}\approx\langle
q_{T}^{2}\rangle.
\end{equation}

Grishchuk's error consists in setting an incorrect initial vacuum
state for the $S$-oscillator, while his choice of the initial
state for the $T$-oscillator is correct. It should be noted that
the vacuum state of the elementary oscillator (\ref{formula-2}) is
unique at stage (\ref{formula-5}) and is determined exclusively by
methods of quantum mechanics\footnote{In essence, it is
mathematics (of Lagrangian systems), or ``\textit{the art of
calling different things by the same names}'', in the definition
by Henry Poincare.}, i.e., knowledge of physics of $T$- and
$S$-modes is not required here. In this sense, all oscillators are
formally similar: their relation to the external field is
determined solely by the function $\alpha(t)$ irrespective of its
physical content (whether it be the scale factor $a$ for the
$T$-oscillator or $a\sqrt{\gamma}$ the $S$-oscillator). In
particular, this means that the amplitude of excitation of the
$S$-oscillator [under the action of the field $\alpha(t)$] from
the minimum-energy state can only depend on the product
$a\sqrt{\gamma}$, and not separately on $\gamma$ or $a$, as was
found in paper \cite{Grishuk} [cf. Eqns (9) and (14)].

In the paper \cite{Grishuk}, the Lagrangian of the $S$-mode
$(22)^{\displaystyle\ast}$ was derived from the Lagrangian of the
$T$-mode $(13)^{\displaystyle\ast}$ with $\tilde{a}\equiv
a\sqrt{\gamma}$ substituted for $a$ and $\bar{\zeta}$ for
$\bar{h}$. However, the correct transition from $T$ to $S$, as
seen from formula (\ref{formula-2}), occurs when substituting
$\alpha_{S}$ for $\alpha_{T}$ and $q_{S}$ for $q_{T}$. Here, to
within a numerical factor of order unity, $a_{0}h\propto \bar h$
goes over into
$\tilde{a}_{0}\zeta\propto\sqrt{\gamma_{0}}\bar{\zeta}$ and not
into $\bar{\zeta}$, as Grishchuk believes. As a result, the
correct Lagrangian (\ref{formula-11}) is obtained by multiplying
Eqn $(22)^{\displaystyle\ast}$ by the factor $\gamma_{0}$.

The Lagrangian $(22)^{\displaystyle\ast}$ is inconsistent with
other formulas from paper \cite{Grishuk}. For example, in the
high-frequency limit gravitational effects are insignificant, and
the Lagrangian for the $S$-mode should turn into the Lagrangian
for sound waves in a medium:
$$
L_{n\gg
aH}=\frac{1}{2n^{3}}a^{2}(\varphi_{1}'^{2}-n^{2}\varphi_{1}^{2})
$$
where
$$
\varphi_{1}\simeq\left(\frac{\gamma}{16\pi
G}\right)^{1/2}\zeta=\frac{\alpha_{S}}{a}q_{S}
$$
is the potential of the matter field for $n\gg aH$; see Eqns
$(19)^{\displaystyle\ast}$-$(20)^{\displaystyle\ast}$. Clearly,
the Lagrangian $(22)^{\displaystyle\ast}$ does not satisfy this
limit.

Another inconsistency: the canonical variables $q_{LPG}$ and
$p_{LPG}$, considered\footnote{Note that the factor $\gamma_{0}$
in Eqns $(24)^{\displaystyle\ast}$ and $(25)^{\displaystyle\ast}$
should be substituted for $\gamma$, since $\zeta$ decreases
inversely proportional to $\tilde{a}=a\sqrt{\gamma}$ in the
adiabatic limit [see Eqn $(20)^{\displaystyle\ast}$].} in Eqns
$(24)^{\displaystyle\ast}$ and $(25)^{\displaystyle\ast}$, are
canonical with regard to Lagrangian (\ref{formula-11}), but not
$(22)^{\displaystyle\ast}$, which is easily verified by directly
inserting expressions (\ref{formula-11}) and
$(22)^{\displaystyle\ast}$ into equation
$(25)^{\displaystyle\ast}$. In addition, by rewriting Lagrangian
$(22)^{\displaystyle\ast}$ in terms of the initial field variable
$\zeta$, we see that it turns to be dependent on the arbitrary
instant of time $\eta_{0}$, which is inadmissible. Removal of the
inconsistency in formula $(22)^{\displaystyle\ast}$ would
eliminate these contradictions.

Summarizing, we can state that Lagrangian
$(22)^{\displaystyle\ast}$ does not follow from the field
Lagrangian for a scalar field minimally coupled with gravity [see
the formula preceding Eqn $(19)^{\displaystyle\ast}$], and then
further discussion is senseless. If one considers Eqn
$(22)^{\displaystyle\ast}$ as a result of a technical inaccuracy
and uses the correct Lagrangian, then the statement on the `false
character' of the standard inflationary result (see, for example,
the title of Section 4 in Ref. \cite{Grishuk}) is due to the
incorrect choice of initial conditions.

Our second remark deals with the measurements of the value of
$T/S$. It is not negligibly small, as the author of paper
\cite{Grishuk} repeatedly states [see, for example, his commentary
to formula $(6)^{\displaystyle\ast}$].

The estimate $T/S\simeq 4\gamma_{i}$ is confirmed by exact
calculations for a broad range of inflationary models (see, for
example, the quantity $r$ in Ref. \cite{Lukash-Mikheeva}). In
particular, all models of the chaotic inflation with $p > 1$
($V(\varphi)\propto\varphi^{2p}$, $p$ is a natural number)
contradict observations since they predict a significant value of
the $T/S$ ratio and a deviation from the Zel'dovich spectrum:
\begin{equation}
\label{formula-15}
\frac{T}{S}\simeq\frac{2p}{N}\simeq(1-n_{S})\frac{2p}{1+p}\approx
0.04p,
\end{equation}
where $N=2\pi G\varphi^{2}/p\approx 50$ on a scale on the order of
$10^{3}$ Mpc.

The case of a massive scalar field ($p=1$) is exceptional in
providing a gravitational-wave mode amplitude only five times
smaller than the scalar one ($\sqrt{0.04}=1/5$), which does not
contradict observational constraints at the 95\% confidence level
(see, for example, Ref. \cite{Seljak}).

It should be emphasized that all values $T/S>0.2$ are excluded by
the modern observations, since in that case the amplitude of the
$S$-mode is insufficient to produce the observed large-scale
structure of the universe (we should remember that the sum $T+S$
is fixed by the data on the anisotropy of cosmic microwave
background).

\section{Acknowledgments}

The author thanks the Editorial Board of \textit{Physics-Uspekhi}
for the invitation to comment on the problem of the relationship
between $T$- and $S$-modes of the primordial cosmological
perturbations. Any discussion of the fundamental problems of
cosmology with such a broad audience as the
\textit{Physics-Uspekhi} readers helps to attract young talented
researchers to this topical problem.

The author is grateful to D.A. Kompaneets and E.V. Mikheeva for
their discussions.

The work is partially supported by the RFBR grant 04-02-17444.


\begin{thebibliography}{99}
\bibitem{Grishuk}
Grishchuk L.P. Usp. Fiz. Nauk \textbf{175} 1289 (2005) [Phys. Usp.
\textbf{49} 1235 (2005)]
\bibitem{Lifshitz}
Lifshitz E.M. Zh. Eksp. Teor. Fiz. \textbf{16} 587 (1946)
\bibitem{Grishuk-2}
Grishchuk L.P. Zh. Eksp. Teor. Fiz. \textbf{67} 825 (1974) [Sov.
Phys. JETP \textbf{40} 409 (1975)]; in Eighth Texas Symp. on
Relativistic Astrophysics (Ann. of the New York Acad. of Sci.,
Vol. 302, Ed.MD Papagian- nis) (New York: The New York Acad. of
Sci., 1977) p. 439
\bibitem{Lukash}
Lukash V.N. Zh. Eksp. Teor. Fiz. \textbf{79} 1601 (1980) [Sov.
Phys. JETP \textbf{52} 807 (1980)]; Pis'ma Zh. Eksp. Teor. Fiz.
\textbf{31} 631 (1980) [JETP Lett. \textbf{31} 596 (1980)];
astro-ph/9910009
\bibitem{Lukash-Mikheeva}
Lukash V.N., Mikheeva E.V. Int. J. Mod. Phys. A \textbf{15} 3783
(2000)
\bibitem{Seljak}
Seljak U. et al. Phys. Rev. D \textbf{71} 103515 (2005)
\end{thebibliography}
\end{document}